\documentclass[%
reprint,
superscriptaddress,
showpacs,preprintnumbers,
 amsmath,amssymb,
 aps,pre,
]{revtex4-1}

\usepackage{graphicx}
\usepackage{dcolumn}
\usepackage{bm}
\usepackage{hyperref}

\begin{document}

\preprint{APS/123-QED}

\title{On the low dimensional dynamics of structured random networks}

\author{Johnatan Aljadeff}
\email{aljadeff@uchicago.edu}
\affiliation{Department of Neurobiology, University of Chicago}
\affiliation{Computational Neurobiology Laboratory, The Salk Institute for Biological Studies}
\author{David Renfrew}%
\affiliation{Department of Mathematics, University of California Los Angeles}%
\author{Marina Vegu\'{e}}
\affiliation{Centre de Recerca Matem\`{a}tica, Campus de Bellaterra, Barcelona, Spain}
\affiliation{Departament de Matem\`{a}tica Aplicada I, Universitat Polit\`{e}cnica de Catalunya, Barcelona, Spain}
\author{Tatyana O. Sharpee}
\affiliation{Computational Neurobiology Laboratory, The Salk Institute for Biological Studies}
\date{\today}

\begin{abstract}
Using a generalized random recurrent neural network model, and by extending our recently developed mean-field approach [J. Aljadeff, M. Stern, T. Sharpee, Phys. Rev. Lett. {\bf 114}, 088101 (2015)], we study the relationship between the network connectivity structure and its low dimensional dynamics. Each connection in the network is a random number with mean 0 and variance that depends on pre- and post-synaptic neurons through a sufficiently smooth function $g$ of their identities.  We find that these networks undergo a phase transition from a silent to a chaotic state at a critical point we derive as a function of $g$. Above the critical point, although unit activation levels are chaotic, their autocorrelation functions are restricted to a low dimensional subspace. This provides a direct link between the network's structure and some of its functional characteristics.  We discuss example applications of the general results to neuroscience where we derive the support of the spectrum of connectivity matrices with heterogeneous and possibly correlated degree distributions, and to ecology where we study the stability of the cascade model for food web structure.

\end{abstract}


\maketitle


\section{Introduction}

Advances in measurement techniques and statistical inference methods allow us to characterize the connectivity properties of large biological systems such as neural and gene regulatory networks \cite{Karlebach2008,Ko2011}. In many cases connectivity is shown to be well modeled by a combination of random and deterministic components. For example, in neural networks, the location of neurons in anatomical or functional space, as well as their cell-type identity influences the likelihood that two neurons are connected \cite{Song2005,Yoshimura2005a,Ko2011}.

For these reasons it has become increasingly popular to study the spectral properties of structured but random connectivity matrices using a range of techniques from mathematics and physics \cite{Sommers1988,Rajan2006,Wei2012,Tao2013,Aljadeff2014a,Ahmadian2015,Aljadeff2015,Muir2015}. In most cases, the spectrum of the random matrix of interest is studied independently of the dynamics of the biological network it implies. Therefore, these results can be used only to make statements about the dynamics of a linear system where knowing the eigenvalues and eigenvectors is sufficient to characterize the dynamics.

Here we study the dynamics of nonlinear random recurrent networks with a continuous synapse-specific gain function that can depend on the pre- and post-synaptic neurons' locations in an anatomical or functional space. These networks become spontaneously active at a critical point that is derived here, directly related to the boundary of the spectrum of a new random matrix model. Given the gain function we predict analytically the network's leading principal components in the space of individual neurons' autocorrelation functions.

In the context of analysis of single and multi-unit recordings our results offer a mechanism for relating structured recurrent connectivity to functional properties of individual neurons in the network; and suggest a natural reduced space where the system's trajectories can be fit by a simple state-space model.

Recently we showed how a certain type of mesoscopic structure can be introduced into the class of random recurrent network models by drawing synaptic weights from a finite number of cell-type-dependent probability distributions \cite{Aljadeff2015}. In contrast to networks with a single cell-type \cite{Sompolinsky1988}, these networks can sustain multiple dynamic global modes.

Here these results are further generalized to networks where the synaptic weight between neurons~$i,j$ is drawn at random from a distribution with mean 0 and variance $N^{-1}g_{ij}^2$, where $N$ is the size of the network.  The smoothness conditions satisfied by the gain function $g$ are stated below. This allows us to treat, for example, networks with continuous spatial modulation of the synaptic gain. The solution to the network's system of mean-field equations that we derive offers a new view-point on how functional properties of single neurons can in fact be a network phenomenon. 

Consider a general synapse-specific gain function $g(z_i,z_j)$ that depends on normalized neuron indices $z_i=i/N$, where $i=1, \dots, N$.  We assume that there is some length scale $s_0>0$ below which $g$ has no discontinuities. That is, we let $g:(0,1]^2\to\mathbb{R}_+$ be a uniformly bounded, continuous function everywhere on the unit square except possibly on a measure zero set $S_0$. The function $g$ may depend on $N$ in such a way that its Lipschitz constant $C_L(N) = C^0_L N^\beta$, with $C^0_L<\infty$ and $1>\beta \ge 0$. Every point where $g$ does not satisfy the above smoothness conditions must be on the boundary between squares of side $s_0$ where it does. 

The network connectivity matrix is then $J\in\mathbb{R}^{N\times N}$ with elements
\begin{equation}
J_{ij} = g(z_i,z_j)J^0_{ij}\label{eq:J}
\end{equation}
where $J_{ij}^0$ is a random matrix with elements drawn at random from a distribution with mean 0, variance $1/N$ and finite fourth moment. In the simulations we use a Gaussian distribution unless noted otherwise. 

In this paper we analyze the the eigenvalue spectrum of the connectivity matrix $J$ and the corresponding dynamics of the neural network. Note that by requiring that $g$ is bounded and differentiable on the unit square outside of $S_0$ we allow the synaptic gain function to be a combination of discrete modulation (e.g. cell-type dependent connectivity for distinct cell-types, as in \cite{Aljadeff2015}) and of continuous modulation (e.g. networks with heterogeneous and possibly correlated in- and out-degree distributions, as in \cite{Roxin2011,Schmeltzer2015}).

When $g$ can be written as an outer product of two vectors (i.e. $g(z_i,z_j) = g_1(z_i)g_2(z_j)$), the model discussed here coincides with that studied by Wei and by Ahmadian et al. \cite{Wei2012,Ahmadian2015}.

The spectral density of $J$ is circularly symmetric in the complex plane, and is supported by a disk centered at the origin with radius $r=\sqrt{\Lambda_1}$ with
\begin{equation}
    \Lambda_1 = {\rm max} \left\{\lambda\left[G_N^{(2)}\right]\right\},
\end{equation}
where $G_N^{(2)}\in \mathbb{R}^{N\times N}_+$ is a deterministic matrix with elements $[G_N^{(2)}]_{ij} = \frac{1}{N}g^2(z_i,z_j)$. Note that $\Lambda_1$ is the Perron-Frobenious eigenvalue of a non-negative matrix, so indeed $\Lambda_1,r\in\mathbb{R}_+$. For general synapse-specific gain function $g$ it has not been possible so far to obtain an explicit formula for $\Lambda_1$. However, we have been able to derive explicit analytic formulae in three cases of biological significance. First, in Section \ref{sec:example} we discuss the case where $G_N^{(2)}$ is a circulant matrix such that $g(z_i,z_j) = g(z_{ij})$ with
\begin{equation}
    z_{ij}= \min\left\{\left|z_i-z_j|,1-|z_i-z_j\right|\right\}
\end{equation}
and show that  $\Lambda_1 = 2\int_0^\frac{1}{2} g^2(z)dz$. This special case is important for large neural networks where connectivity often varies smoothly as a function of neuron's index. In Section \ref{sec:degdist} we derive the support of the bulk spectrum and the outliers of a random connectivity matrix with heterogeneous joint in- and out-degree distribution. Finally, in Section \ref{sec:ecology} we discuss a third example pertinent to large scale models of ecosystems. These systems are often modeled using  $g$ that has a triangular structure and again there is an analytic formula for $\Lambda_1$ in this case.

Given the connectivity matrix $J$ defined in Eq.~(\ref{eq:J}), the dynamics of neural network model with $N$ neurons is described by
\begin{equation}
    \dot{x}_i(t) = -x_i(t) + \sum_{j=1}^N J_{ij} \phi_j(t),\label{eq:dyn0}
\end{equation}
where $\phi_j(t) = \tanh[x_j(t)]$. The $x$ variables can be thought of as the membrane potential of each neuron, and the $\phi$ variables as the deviation of the firing rates from their average values.

Using a modified version of dynamic mean field theory we show that in the limit $N\to\infty$ this system undergoes a phase transition, where $r$ is the coordinate that describes this transition and $r=1$ is the critical point. Below the critical point ($r<1$), the neural network has a single stable fixed point at $x=0$. Above the critical point the system is chaotic.

We analyze the dynamics above the critical point in more detail and find a direct link between the network structure ($g$) and its functional properties. To that end we define $N$ dimensional autocorrelation vectors
\begin{equation}
\Delta_i(\tau) = \left\langle x_i(t)x_i(t+\tau) \right\rangle,~~C_i(\tau) = \left\langle \phi_i(t)\phi_i(t+\tau) \right\rangle\label{eq:aci}
\end{equation}
where $\langle\cdot\rangle$ denotes average over the ensemble of matrices $J$ and time. These vectors are restricted to the potentially low dimensional subspace spanned by the right eigenvectors of $G_N^{(2)}$ with corresponding eigenvalues that have real part greater than $1$. Thus, although the network dynamics are chaotic, they are confined to a low dimensional space, which has been suggested as a mechanism that could make computation in the network more robust \cite{Sussillo2013}. 

\section{Derivation of the critical point}\label{sec:crit}

\subsection{Finite number of partitions}
We begin by recalling our recent results for a function $g$ that has block structure. We defined a $D\times D$ matrix with elements $g_{cd}$ and partitioned the indices $1,\dots,N$ into $D$ groups, with the $c$-th partition have a fraction $\alpha_c$ neurons. The synaptic gain function was then defined by $g(z_i,z_j)=g_{c_i c_j}$, where $c_i$ is the partition index of the $i$-th neuron.

Defining $n_d = N\sum_{c=1}^d\alpha_c$ allows us to write formally $c_i = \left\{ c \left| i \in \left(n_{c-1}, n_c\right]\right.\right\}$. With these definitions, we rewrite Eq.~(\ref{eq:dyn0}) in a form that emphasizes the separate contributions from each group to a neuron:
\begin{equation}
    \dot{x}_i = -x_i + \sum_{d = 1}^D g_{c_i d}\sum_{j = n_{d - 1} + 1}^{n_d}J^0_{ij}\phi_j\left(t\right).\label{eq:dyn}
\end{equation}

In \cite{Aljadeff2015} we used the dynamic mean field approach \cite{Amari1972,Sompolinsky1982,Sompolinsky1988} to study the network behavior in the $N\rightarrow\infty$ limit. Averaging Eq.~(\ref{eq:dyn}) over the ensemble from which $J$ is drawn implies that neurons that belong to the same group are statistically identical. Therefore, to represent the network behavior it is enough to look at the activities $\xi_d(t)$ of $D$ representative neurons and their inputs $\eta_d\left(t\right)$.

The stochastic mean field variables $\xi(t)$ and $\eta(t)$ will approximate the activities and inputs in the full $N$ dimensional network provided that they satisfy the dynamic equation
\begin{equation}
\dot{\xi}_d\left(t\right) = -\xi_d\left(t\right)+\eta_d\left(t\right),
\label{eq:mfield}
\end{equation}
and provided that $\eta_d\left(t\right)$ is drawn from a Gaussian distribution with moments satisfying the following conditions. First, the mean $\langle\eta_d(t)\rangle = 0$ for all $d$. Second, the correlations of $\eta$ should match the input correlations in the full network, averaged separately over each group. Using Eq. (\ref{eq:mfield}) and the property $N\left\langle J^0_{ij}J^0_{kl}\right\rangle = \delta_{ik} \delta_{jl}$ we get the self-consistency conditions:
\begin{eqnarray}
\left\langle\eta_c\left(t\right)\eta_d\left(t+\tau\right)\right\rangle & =& \delta_{cd}\sum_{b=1}^D\alpha_b g^2_{cb}C_b(\tau),
\label{eq:mf_constraints}
\end{eqnarray}
where $\left\langle \cdot \right\rangle$ denotes averages over $i = n_{c-1}+1,\dots,n_c$ and $k = n_{d-1}+1,\dots,n_d$ in addition to average over realizations of $J$. The average firing rate correlation vector is denoted by $C\left(\tau\right)$. Its components (using the variables of the full network) are
\begin{equation}
    C_d(\tau)=\frac{1}{N\alpha_d}\sum^{n_d}_{i=n_{d-1}+1} \left\langle\phi[x_i(t)]\phi[x_i(t+\tau)]\right\rangle,
\end{equation}
translating to
\begin{equation}
    C_d (\tau) = \left\langle\phi[\xi_d(t)]\phi[\xi_d(t+\tau)]\right\rangle
\end{equation}
using the mean field variables. Importantly, the covariance matrix $\mathcal{H}(\tau)$ with elements $\mathcal{H}_{cd}\left(\tau\right)=\left\langle\eta_c\left(t\right)\eta_d\left(t+\tau\right)\right\rangle$ is diagonal, justifying the definition of the vector $H=\text{diag}\left(\mathcal{H}\right)$. With this in hand we rewrite Eq.~(\ref{eq:mf_constraints}) in matrix form as
\begin{equation}
    H\left(\tau\right) = MC\left(\tau\right),
    \label{eq:matrix_form}
\end{equation}
where $M\in\mathbb{R}_+^{D\times D}$ is a constant matrix reflecting the network connectivity structure: $M_{cd} = \alpha_d g^2_{cd}$.

A trivial solution to this equation is $H(\tau)=C(\tau)=0$ which corresponds to the silent network state: $x_i(t) = 0$. Recall that in the network with a Girko matrix as its connectivity matrix ($D=1$), the matrix $M = g^2$ is a scalar and Eq.~(\ref{eq:matrix_form}) reduces to $H(\tau) = g^2 C(\tau)$. In this case the silent solution is stable only when $g<1$. For $g>1$ the autocorrelations of $\eta$ are non-zero which leads to chaotic dynamics in the $N$ dimensional system \cite{Sompolinsky1988}.

When $D>1$, Eq.~(\ref{eq:matrix_form}) can be projected on the eigenvectors of $M$ leading to $D$ consistency conditions, each equivalent to the single group case. Each projection has an effective scalar given by the eigenvalue in place of $g^2$ in the $D=1$ case. Hence, the trivial solution will be stable if all eigenvalues of $M$ have real part $<1$. This is guaranteed if $\Lambda_1$, the largest eigenvalue of $M$, is $<1$. If $\Lambda_1>1$ the projection of Eq.~(\ref{eq:matrix_form}) on the leading eigenvector of $M$ gives a scalar self-consistency equation analogous to the $D=1$ case for which the trivial solution is unstable. As we know from the analysis of the $D=1$ case, this leads to chaotic dynamics in the full network. Therefore $\Lambda_1=1$ is the critical point of the $D>1$ network. Furthermore, the fact that in the $D=1$ case the presence of the destabilized fixed point at $x=0$ corresponds to a finite mass of the spectral density of $J$ with real part $>1$ \cite{Sompolinsky1988,Molgedey1992} allowed us to read the radius of the support of the connectivity matrix with $D>1$ and identify it as $r=\sqrt{\Lambda_1}$ \cite{Aljadeff2015}.

\subsection{The continuous case}
The vector dynamic mean field theory we developed in \cite{Aljadeff2015} relies on having an infinite number of neurons in each partition with the same statistics. The natural choice is therefore to have the size of each group of neurons be linear in the system size: $N_c = \alpha_c N$.

This scaling imposes two limitations if one wishes to compare the results to the dynamics of more realistic networks. It requires knowledge of the cell-type-identity of each neuron in the recording, which often is not available; and it confines the statements we are able to make about the dynamics to quantities that are averaged over neurons that belong to the same cell-type.

To lift the requirement of block structured variances (i.e. now $g=g(z_i,z_j)$), we can do the following. Let $K(N)\in\mathbb{N}$ be a weakly monotonic function of $N$ such that
\begin{equation}
    \lim_{N\to\infty} \frac{K(N)}{N} =  0, \quad \lim_{N\to\infty} \frac{N^\beta}{K(N)} = 0.
\end{equation}
Recall that $1>\beta\ge 0$ and that the Lipschitz constant of $g$ scales as $N^\beta$, implying that $\lim_{N\to\infty} K(N) =  \infty$. A natural choice is $K(N) = N^{\tilde{\beta}}$ with $\tilde{\beta} = 1-\beta$, but as long as $1>\tilde{\beta}>\beta$ the specific scaling behavior will not matter in our analysis. For convenience we will suppress the $N$ dependence when possible.

Let $\mu = 1,\dots,K$ and let 
\begin{equation}
\mu_i = \left\{ \mu \left| \frac{i}{N} \in \left(\frac{\mu-1}{K},\frac{\mu}{K}\right]\right.\right\}.\label{eq:mui}
\end{equation} 
Furthermore, define $\tilde{g}\in \mathbb{R}_+^{N\times N}$ with elements
\begin{eqnarray}
\tilde{g}_{ij} & = &  g\left(\frac{\mu_i-\frac{1}{2}}{K},\frac{\mu_j-\frac{1}{2}}{K}\right).\label{eq:piecewise}
\end{eqnarray}
In other words, $\tilde{g}$ is an $N\times N$ matrix with $K^2$ equally sized square blocks. The value of elements in each block is the value of the function $g$ in the middle of that block. These definitions allow us to bridge the gap between the block and the continuous cases for the following reasons. Consider the random connectivity matrix with elements $\tilde{J}_{ij} = \tilde{g}_{ij}J^0_{ij}$ and the network that has $\tilde{J}$ as its connectivity.

First, since $N/K\to \infty$ as $N\to\infty$, the number of neurons in each group goes to infinity, and we may use the vector dynamic mean field theory as before, but in a $K$ dimensional space (rather than $D$ which was $O(1)$).

The critical point is now given in terms of the largest eigenvalue of an $N\times N$ matrix $\tilde{M}$ with elements
\begin{equation}
    \tilde{M}_{ij} = \frac{1}{N}g^2 \left(\frac{\mu_i-\frac{1}{2}}{K},\frac{\mu_j-\frac{1}{2}}{K}\right).
\end{equation}
where ${\rm rank}\{ \tilde{M}\}\le K$. 

Second, recall that the partitioning of the matrix $\tilde{g}$ depends on $N$ and the function $g$ is assumed to be smooth outside of a set with measure zero $S_0$. These properties will allow us to show (see Appendix \ref{sec:KN}) that as $N\to\infty$ we have
\begin{equation}
\tilde{g}_{ij} \to g(z_i,z_j),\qquad \tilde{M}_{ij} \to \left[G^{(2)}_N\right]_{ij},
\end{equation}
meaning that by studying the system with connectivity structure $\tilde{g}$ in the limit $N\to\infty$ we are in fact obtaining results for the generalized connectivity matrix with a smooth synaptic gain function $g$.

\subsection{Circular symmetry of spectrum}

In \cite{Aljadeff2014a} we used random matrix theory techniques to derive, for the case of block-structured $J$, an implicit equation that the full spectral density of $J$ satisfies. The circular symmetry of the spectrum for that case is obvious because the equations (see Eq. 3.6 in \cite{Aljadeff2014a}) depend on the complex variable $z$ only through $|z|^2$. Similar implicit equations, with integrals instead of sums, can be written for the continuous case. Rigorous mathematical analysis of the spectral density implied by such equations is beyond the scope of this paper and will be presented elsewhere. Nevertheless, the integral equations still depend on $|z|^2$, supporting the circular symmetry of the spectrum.

\begin{figure*}
\begin{centering}
    \includegraphics[width=0.8\textwidth]{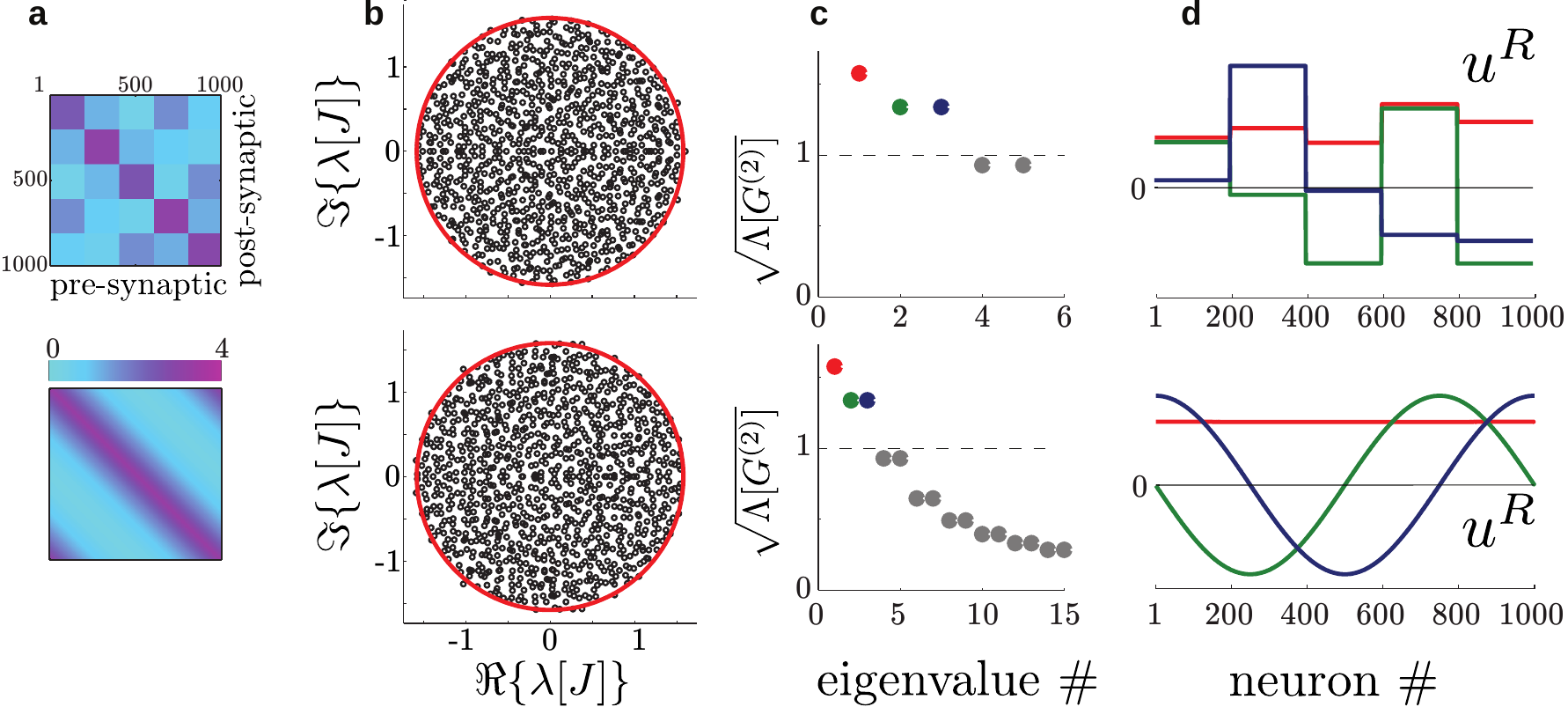}
    \caption{Eigenspaces of two example networks - one with block structured connectivity (top) and another with continuous gain modulation (bottom). (a) The synaptic gain matrix $g_{ij}$. (b) The spectrum of the random connectivity matrix $J$ in the complex plane. The spectrum is supported by a disk with radius $r=\sqrt{\Lambda_1}$ indicated in red. (c) The square root of the largest eigenvalues of $G^{(2)}_N$. When these are greater than 1, the corresponding eigenvectors (shown in (d)) are active autocorrelation modes. For the continuous function we chose the circulant parametrization (see Section \ref{sec:gcirc}) with $g_0=0.3$, $g_1=3.0$ and $\gamma=2.0$. For the block structured connectivity, $g$ was chosen such that the first 5 eigenvalues match exactly to those of the continuous network.}
    \label{fig:spectrum}
\end{centering}
\end{figure*}

\section{Dynamics above the critical point}\label{sec:dyn}

\subsection{Finite number of partitions}

To study the spontaneous dynamics above the critical point we recall again the analogous result for a matrix with block structure. The $D$ dimensional average autocorrelation vectors $C(\tau),~\Delta(\tau)$ (see definition below) are restricted to a $D^\star$ dimensional subspace, where $D^\star$ is the number of eigenvalues of $M$ with real part $>1$ (i.e. the algebraic multiplicity of these eigenvalues). This result is obtained by projecting Eq. (\ref{eq:matrix_form}) on the Schur basis vectors of $M$ \cite{Aljadeff2015}.

The definitions of the $d=1,\dots,D$ component of these vectors are
\begin{eqnarray}
\Delta_d(\tau) &=& \frac{1}{N\alpha_d}\sum_{i=n_{d-1}+1}^{n_d} \langle x_i(t)x_i(t+\tau) \rangle \\
C_d(\tau) &=& \frac{1}{N\alpha_d}\sum_{i=n_{d-1}+1}^{n_d} \langle \phi_i(t)\phi_i(t+\tau) \rangle,
\end{eqnarray}
and the $D^\star$ dimensional subspace is
\begin{equation}
U_M = {\rm span}\{u_1^R,\dots,u_{D^\star}^R\} \label{eq:UM}
\end{equation}
where $u^R_d$ are the right eigenvectors of $M$ in descending order of the real part of their corresponding eigenvalue (see examples in Fig. \ref{fig:spectrum}). An equivalent statement is that, independent of the lag $\tau$, projections of the vectors $C(\tau),~\Delta(\tau)$ on any vector in the orthogonal complement subspace $U^\perp_M$ are approximately $0$. Note that for asymmetric (but diagonalizable) $M$, $U^\perp_M$ is spanned by the left rather than the right eigenvectors of $M$: 
\begin{equation} 
U^\perp_M = {\rm span}\{u_{D^\star+1}^L,\dots,u_{D}^L\}.\label{eq:UMperp}
\end{equation}

\subsection{Autocorrelation modes in the generalized model}

We can repeat the analysis of \cite{Aljadeff2015} for a network with connectivity $\tilde{J} = \tilde{g}_{ij}J^0_{ij}$ that has $K^2$ blocks, and for each $N,K(N)$ obtain the subspace $U_{\tilde{M}}$ that the $K$ dimensional autocorrelation vectors $\tilde{C}(\tau),~\tilde{\Delta}(\tau)$ are restricted to. These vectors have components
\begin{eqnarray}
\tilde{\Delta}_\mu(\tau) & = & \frac{1}{K}\sum_{i=\frac{N}{K}(\mu-1)+1}^{\frac{N}{K}\mu} \langle x_i(t)x_i(t+\tau) \rangle \\
\tilde{C}_\mu(\tau) & = & \frac{1}{K}\sum_{i=\frac{N}{K}(\mu-1)+1}^{\frac{N}{K}\mu} \langle \phi_i(t)\phi_i(t+\tau) \rangle.
\end{eqnarray}

Now when we take the limit $N\to\infty$ the dimensionality of the autocorrelation vectors $\tilde{C}(\tau),~\tilde{\Delta}(\tau)$ becomes infinite as well, but the subspace $U_{\tilde{M}}$ may be of finite dimension $K^\star$, where $K^\star$ is the algebraic multiplicity of eigenvalues of $\tilde{M}$ with real part greater than 1 (see Section \ref{sec:example} for an example).

We have shown that for $g$ that satisfies the smoothness conditions, studying the network with connectivity $J_{ij} = g(z_i,z_j)J^0_{ij}$ is equivalent to studying the network with connectivity $\tilde{J}$ in the limit $N\rightarrow\infty$. Therefore, in that limit, the individual neuron autocorrelation functions $C_i(\tau),~\Delta_i(\tau)$ (Eq. \ref{eq:aci}) are restricted to the subspace spanned by the right eigenvectors of $\tilde{M} \to G_N^{(2)}$ corresponding to eigenvalues with real part $>1$.

\begin{figure*}
\begin{centering}
    	\includegraphics[width=0.75\textwidth]{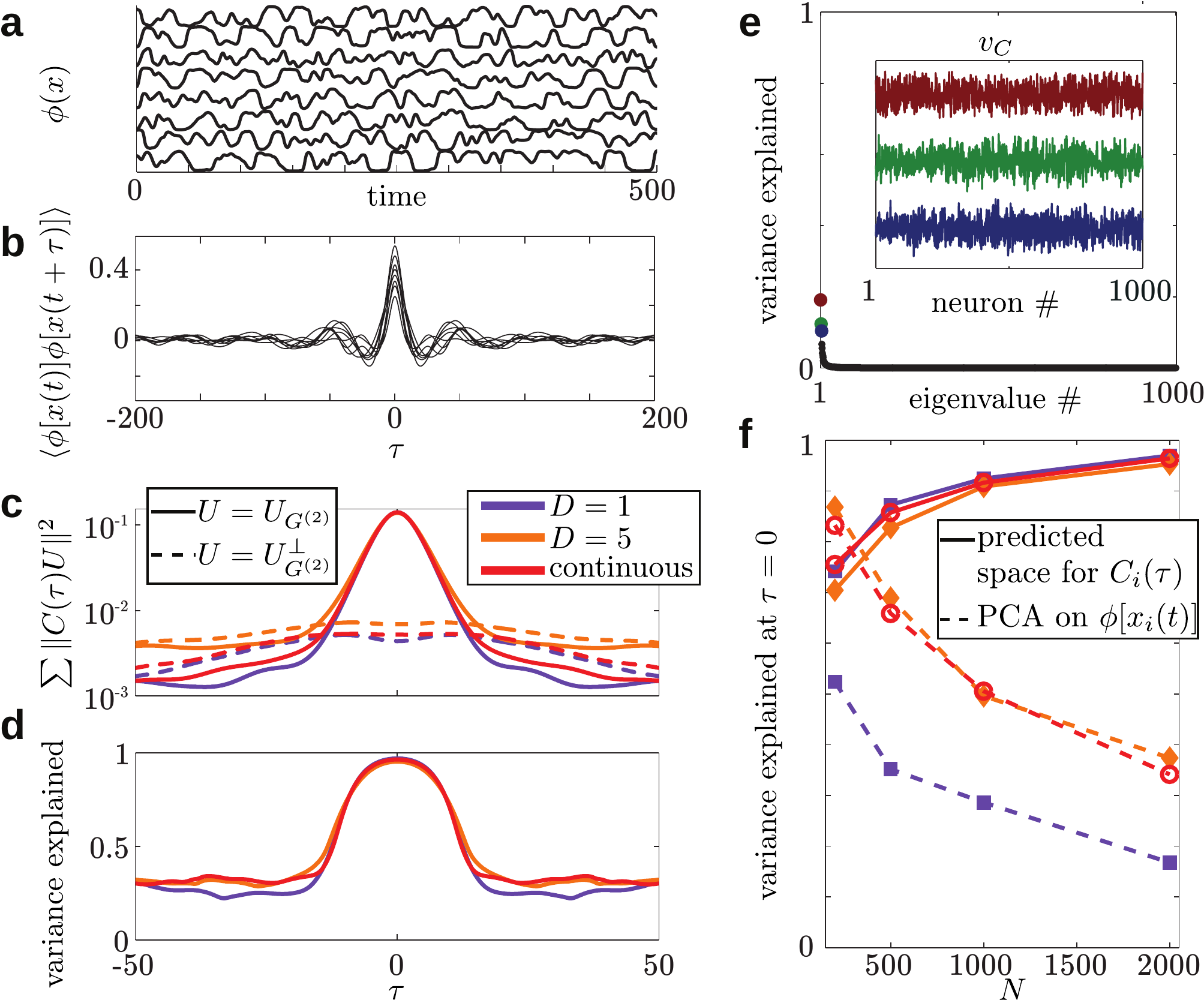}
    \caption{Low dimensional structure of network dynamics. Traces of the firing rates $\phi[x_i(t)]$ (a) and autocorrelations $C_i(\tau)$ (b) of eight example neurons chosen at random from the network with continuous gain modulation (shown in the bottom row of Fig. \ref{fig:spectrum}). (c) The sum of squared projections of the vector $C_i(\tau)$ on all active and inactive autocorrelation modes (solid and dashed lines, respectively). The dimension of the subspace $U_{G^{(2)}}$ is $K^\star=1$ for the network with $g=\text{const}$ and $K^\star=3$ for the block and continuous cases (orange and red, respectively), much smaller than $N-K^\star\approx N$, the dimension of the orthogonal complement space $U_{G^{(2)}}^\perp$. (d) Our analytically derived subspace accounts for almost 100 percent of the variance in the autocorrelation vector for $|\tau|\lesssim 10$ (in units of the synaptic time constant). (e) Reducing the dimensionality of the dynamics via Principal Component Analysis on $\phi(x)$ leads to vectors (inset) that account for a much smaller portion of the variance (when using same dimension $K^\star$ for the subspace), and lack structure that could be related to the connectivity. (f) Summary data from 50 simulated networks per parameter set ($N$, structure type) at $\tau=0$. As $N$ grows the \emph{leak} into $U_{G^{(2)}}^\perp$ diminishes if one reduces the space of the $C_i(\tau)$ data while the fraction of variance explained becomes smaller when using PCA on the $\phi[x_i(t)]$ data, a signature of the extensiveness of the dimension of the chaotic attractor.}
    \label{fig:ac}
    
\end{centering}
\end{figure*}

This in fact is equivalent to, given the network structure $g$, predicting analytically the leading principal components in the $N$ dimensional space of individual neuron autocorrelation functions (see Fig. \ref{fig:ac}). Note that traditionally principal component analysis is performed in the $N$ dimensional space of neuron firing rates rather than autocorrelation functions. Numerical analysis performed in \cite{Rajan2010a} suggests that the system's trajectories, when considered in the space spanned by the vectors $x$ or $\phi(x)$ (individual neuron activations/firing-rates), occupy a space of dimension that is extensive in the system size $N$. However, when considered in the space of individual neuron autocorrelation functions, the dimension of trajectories is intensive in $N$ and usually finite. In the subspace we derive here the information about the relative phases between neurons is lost, but the amplitude and frequency information is preserved. Section \ref{sec:conc} includes further discussion of the consequences of our predictions and how they may be applied.

\subsection{Finite $N$ behavior}

For a finite system it is evident from numerical simulations that the $N$ dimensional vector of autocorrelation functions ``leaks'': it has non-zero  projections on inactive modes - eigenvectors of $G_N^{(2)}$ with corresponding eigenvalue which is $<1$ (see Fig. \ref{fig:ac}). Here we study the magnitude of this effect, and specifically its dependence on $N$ and on the model's structure function $g$. For simplicity, we will study the projections of the autocorrelation vector $C(\tau)$ at lag $\tau=0$. Let
\begin{equation}
\sigma^2_C(g,N) = \left\langle \left\|C^\top(0)U^\perp(g,N) \right\|^2 \right\rangle
\end{equation}
where $U^\perp(g,N)$ is an $N\times (N-K^\star)$ matrix with columns equal to orthogonalized eigenvectors of $G_N^{(2)}$ (i.e. the Schur basis vectors) with corresponding eigenvalue less than 1, see Eqs. (\ref{eq:UM}) and (\ref{eq:UMperp}). Here $\langle\cdot\rangle$ denotes averaging over an ensemble of connectivity matrices (with the same structure $g$ and same size $N$).

Consider the homogeneous network (i.e. constant $g(z_i,z_j)=g_0>1$). Now $U^\perp(g,N)$ contains all the vectors in $\mathbb{R}^N$ perpendicular to the DC mode $\frac{1}{\sqrt{N}}[1,\dots,1]$. Thus, $\sigma^2_C(g_0,N)$ is simply the variance over the neural population of the individual neuron autocorrelation functions at lag $\tau=0$.

We can now use the mean-field approximation to determine the $N$ dependence of $\sigma^2_C(g_0,N)$. For $N\gg 1$, the elements of the vector $C(0)$ follow a scaled $\chi^2$ distribution
\begin{equation}
    C_i(0) = N^{-1} q(g_0^2) y^N_i,\qquad y^N_i  \sim \chi^2(N),
\end{equation}
where $q(g_0)\sim\mathcal{O}(g_0)\sim\mathcal{O}(1)$ and $\chi^2(N)$ is the standard chi-squared distribution with $N$ degrees of freedom.
Thus, in this limit,
\begin{eqnarray}
    \langle C_i(0)\rangle & = & q(g_0^2),\\
    \langle C_i(0)C_j(0)\rangle-\langle C_i(0)\rangle\langle C_j(0)\rangle   & = & 2\delta_{ij}N^{-1}q^2\left(g_0^2\right)\approx \delta_{ij}/N. \nonumber
\end{eqnarray}
The autocorrelation function is in general a single neuron property. Therefore, their variation about the mean is uncorrelated across neurons independent of the network's structure: $\langle C_i(0)C_j(0)\rangle-\langle C_i(0)\rangle\langle C_j(0)\rangle\propto\delta_{ij}$. Thus, we can use the notation $\langle(\delta C_i(0))^2\rangle = \langle C_i(0)C_i(0)\rangle-\langle C_i(0)\rangle\langle C_i(0)\rangle$.

Similarly, in the case with $D$ partitions,
\begin{eqnarray}
    \langle C_i(0)\rangle & =& q_{c_i}(M),\nonumber\\
    \langle(\delta C_i(0))^2\rangle &=& 2q_{c_i}^2(M)(\alpha_{c_i}N)^{-1}\approx D/N.\label{eq:varCD}
\end{eqnarray}
Finally, for $K(N)$ partitions,
\begin{eqnarray}
    \langle C_i(0)\rangle & = & q_i(g_N^{(2)}),\nonumber\\
    \langle(\delta C_i(0))^2\rangle & =& 2q_i^2(G_N^{(2)})(K/N) \approx K[g]/N.\label{eq:varCK}
\end{eqnarray}
At this stage, Eq. (\ref{eq:varCK}) remains ambiguous because the function $K(N)$ is not a property of the neural network model. Rather, it is a construction we use to show that in the limit $N\rightarrow\infty$ we are able to characterize the dynamics using the vector dynamic mean field approach. Therefore, for finite $N$ we now wish to estimate an appropriate value of $K=K[g]$.

This can be done by noting that the network with block structured connectivity is a special case of the one with a continuous structure function. For that special case we know that $K[g]=D$. Since $g$ is smooth, for sufficiently large $K_0$, we can assume that in each block $g$ is linear in both variables $z_i$ and $z_j$:
\begin{eqnarray}
    g(z_i,z_j) & \approx & \tilde{g}_{ij} + \tilde{g}_N^{(1,0)}(\mu_i,\mu_j)\left(z_i-\frac{\mu_i-\frac{1}{2}}{K}\right) + \nonumber \\
               & &  \tilde{g}_N^{(0,1)}(\mu_i,\mu_j)\left(z_j-\frac{\mu_j-\frac{1}{2}}{K}\right).
\end{eqnarray}
Here $\tilde{g}_N^{(1,0)}(\mu_i,\mu_j)$ is the first derivative of $g$ with respect to the first variable, evaluated in the middle of the $\mu_i,\mu_j$ block.

The only expression for $K[g]$ that depends on first derivatives of $g$ and agrees with the homogeneous and block cases is
\begin{eqnarray}
    K[g] &\approx & 1 + \iint \left[\left|g^{(1,0)}(x,y)\right|+\left|g^{(0,1)}(x,y)\right|\right]dxdy\nonumber \\
         & \approx & 1 + \iint \|\nabla g\|dxdy.
\end{eqnarray}

We are unable to test this prediction quantitatively, because we do not know the dependence of the function $q$ on the structure $g$. We are able to show however that the dependence on $N$ is the same as for the block models, which is confirmed by numerical simulations (compare solid purple, orange and red lines in Fig. \ref{fig:ac}f). In the cases where $g$ depends on $N$, the value of $K[g]$ will also depend on $N$, such that the scaling of the ``leak'' may no longer be $\propto N^{-1}$.  

\section{An example where $g$ is circulant}\label{sec:example}

When the matrix $g(z_i,z_j)$ is circulant such that $g(z_i,z_j) = g(z_{ij})$ with
\begin{equation}
    z_{ij}= \min\left\{\left|z_i-z_j|,1-|z_i-z_j\right|\right\},
\end{equation}
the eigenvalues and eigenvectors of $G^{(2)}_N$ are given in closed form by integrals of the function $g^2(z_{ij})$ and the Fourier modes with increasing frequency. In particular, the largest eigenvalue $\Lambda_1 = 2\int_0^\frac{1}{2}g^2(z)dz$ corresponds to the zero frequency eigenvector $\propto[1,\dots,1]$. To show this, consider the $k+1$ eigenvalue of the circulant matrix $G_N^{(2)}$:
\begin{equation}
\Lambda_{k+1} = \frac{1}{N}\sum_{j=1}^N \exp\left(\frac{2\pi ijk}{N}\right)g^2(z_1,z_j).
\end{equation}
So in the limit $N\to\infty$,
\begin{eqnarray}
\Lambda_{k+1} &=&  \lim_{N\to\infty}\frac{2}{N}\sum_{j=1}^{N/2} \exp\left(2\pi ikz_{1j}\right)g^2(z_{1j}) \nonumber \\ 
&=& 2\int_0^\frac{1}{2} \exp\left(2\pi ikz\right)g^2(z)dz,
\end{eqnarray}
as desired.

\subsection{A ring network} \label{sec:gcirc}

As an example we study a network with ring structure that will be defined by $g(z_i,z_j) = g_0 + g_1 (1-2z_{ij})^\gamma$, such that neurons that are closer are more strongly connected.

This definition leads to the following form for the critical coordinate along which the network undergoes a phase transition
\begin{equation}
\Lambda_1 = g_0^2 +  \frac{2g_0g_1}{\gamma+1} + \frac{g^2_1}{2\gamma+1}
\end{equation}
Interestingly, as $g_1$ increases continuously, additional discrete modes with increasing frequency over the network's spatial coordinate become active by crossing the critical point $\Lambda_k=1$. When modes with sufficiently high spatial frequency have been introduced, nearby neurons may have distinct firing properties.

\subsection{A toroidal network}

In contrast to the ring network discussed above, the connectivity in real networks often depends on multiple factors. These could be the spatial coordinates of the cell body or the location in a functional space (e.g. the frequency that each particular neuron is sensitive to). Therefore we would like to consider a network where the function $g$ depends on the distance between neurons embedded in a multidimensional space.

This problem was recently addressed by Muir and Mrsic-Flogel \cite{Muir2015} by studying the spectrum of a specific type of Euclidean random matrix. In their model, neurons were randomly and uniformly distributed in a space of arbitrary dimension, and the connectivity was a deterministic function of their distance. While their approach resolves the issue of the spectral properties of the random matrix when connectivity depends on distance in more than one dimension, the dynamics these matrices imply remain unknown.
 
\begin{figure*}
\begin{centering}
    \includegraphics[width=1\textwidth]{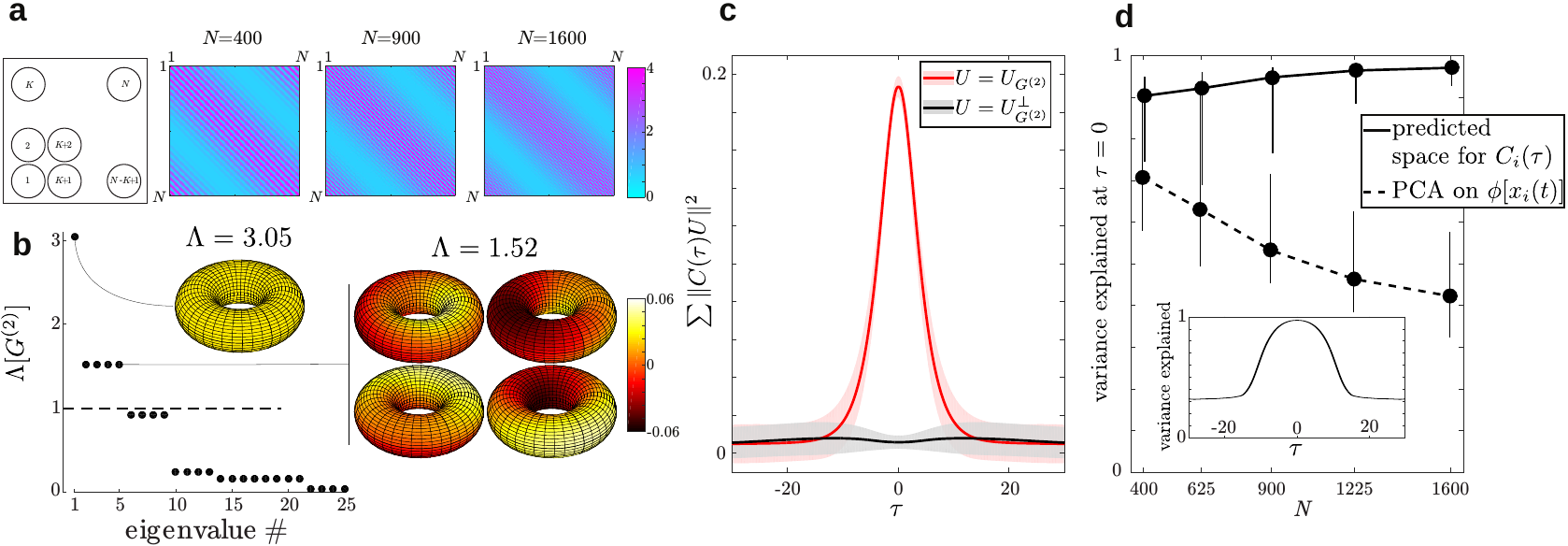}
    \caption{(a) A grid strategy with $K=\sqrt{N}$ for tiling the $[0,1] \times [0,1]$ torus with $N$ neurons (left) and the resulting deterministic gain matrix with elements $g_{ij}$ for three values of $N$ as defined in Eq. (\ref{eq:torusg}) (right). Unlike the ring network, here $g$ depends on $N$, and its derivative is unbounded so as $N$ increases the gain function ``folds''. The parameters of the connectivity matrix are
 $g_0=0.7$, $g_1=0.8$. (b) The 25 non-zero eigenvalues of $G_N^{(2)}$ for $N=1600$ and the eigenvectors corresponding to eigenvalues that are greater than 1 plotted on a torus with coordinates $(\theta_i^1,\theta_i^2)$. (c) The sum of squared projections of the vector $C_i(\tau)$ on all active and inactive autocorrelation modes (red and black lines, respectively). Shades indicate the standard deviation computed from 50 realizations. (d) Comparison of the variance explained at $\tau=0$ by our predicted subspace (solid line) and by performing PCA on $\phi(x)$ (dashed line). Error bars represent 95\% confidence intervals. Inset: the subspace we derived accounts for a large portion of the variance for time lags $|\tau|\lesssim 10$ (in units of the synaptic time constant).}
    \label{fig:torus}
\end{centering}
\end{figure*}

To study the spectrum and the dynamics jointly, we define a network where neurons' positions form a square $K\times K$ grid (with $K=\sqrt{N}$) on the $[0,1] \times [0,1]$ torus (see Fig. \ref{fig:torus}a):
\begin{equation}
\theta^1_i = \frac{\left\lfloor i/K\right\rfloor}{K},\qquad \theta^2_i = \frac{i~{\rm mod}~K}{K}.\label{eq:toruscoordinates}
\end{equation}
The positions of the neurons on the torus are schematized in Fig. \ref{fig:torus}a.

An analogue parameterization for $g$ to the one we used in the ring example which respects the toroidal geometry reads
\begin{equation}
g_{ij}  = g_0+g_1 \Big[\cos\big(2\pi z_{ij}\big)+1\Big]\Big[\cos\big(2\pi\sqrt{N}z_{ij}\big)+1\Big].\label{eq:torusg}
\end{equation}
Note that now $g$ depends on $N$, but it is bounded and its Lipschitz constant scales as $\sqrt{N}$, so it satisfies the smoothness conditions. 

Fig. \ref{fig:torus}b shows the spectrum of $G^{(2)}$ and the corresponding eigenvectors, plotted on a torus. Because there are non-uniform modes that are active (two through five), then each neuron has a different participation in the vector of autocorrelation functions. In Fig. \ref{fig:torus}c,d we show for a network with a range of $N$ values that indeed the vector of autocorrelation functions is restricted to the predicted subspace in contrast to the firing rate vector.

The gain function analyzed here depends on a Euclidean distance on the torus. Other metrics, for example a city-block norm, can be treated similarly. 

Overall these results provide a mechanism whereby continuous and non-fine tuned connectivity that depends on a single or multiple factors can lead to a few active dynamic modes in the network. Importantly, the modes maintained by the network inherit their structure from the deterministic part of the connectivity.

\section{Matrices with heterogeneous degree distributions}\label{sec:degdist}

Here we will use our general result to compute the spectrum of a random connectivity matrix with specified in- and out-degree distributions. Realistic connectivity matrices found in many biological systems have degree distributions which are far from the binomial distribution that would be expected for standard Erd\H{o}s-R\'{e}nyi networks \cite{Barabasi1999}. Specifically, they often exhibit correlation between the in- and out-degrees, clustering, community structures and possibly heavy-tailed degree distributions \cite{Song2005,Boccaletti2006}. We consider a connectivity matrix appropriate for a neural network model. Since each element of this matrix will have a non-zero mean, our current theory cannot make statements about the dynamics. Nevertheless the spectrum of the connectivity matrix is important on its own as a step towards understanding the behavior of random networks with general and possibly correlated degree distributions.

Consider a network with $N_E$ excitatory and $N_I$ inhibitory neurons ($N=N_E+N_I$). Each inhibitory neuron has incoming and outgoing connections with probability $p_0$ to and from every other neuron in the network. Within the excitatory subnetwork, degree distributions are heterogeneous. Specifically, $k^{\text{in}},k^{\text{out}}$ are the average excitatory in- and out-degree sequences that are drawn from a joint degree distribution that could be correlated. We assume that $\sum_{i=1}^{N_E} k_i^{\text{in}}=\sum_{i=1}^{N_E} k_i^{\text{out}}=N_E \bar{k}$,  where $\bar k$ is the mean connectivity, and that the marginals of the degree distribution are equal. Define $x,y$ to be the $N_E$ dimensional vectors  $x=k^{\text{in}}/\sqrt{N_E \bar{k}}$ and $y=k^{\text{out}}/\sqrt{N_E \bar{k}}$.

The matrix $P$ defines the probability of connections given the fixed normalized degree sequences and $p_0$:
\begin{eqnarray}
P_{ij} & = & \begin{cases}
x_iy_j & 1\le i,j\le N_E\\
p_0 & \text{otherwise}
\end{cases}.
\end{eqnarray}
The random adjacency matrix is then $A_{ij}\sim\text{Bernoulli}\left(P_{ij}\right)$. Note that because the adjacency matrix is random, $k^{\text{in}}$ and $k^{\text{out}}$ are the \emph{average} in- and out-degree sequences. 

The connectivity matrix is then
\begin{eqnarray}
J_{ij} & = & A_{ij}W_{ij}
\end{eqnarray}
with 
\begin{eqnarray}
W_{ij} & = & \begin{cases}
-W_0 & j>N_E\\
1 & \text{otherwise}
\end{cases},
\end{eqnarray}
where $W_0$ is the ratio of the synaptic weight of inhibitory to excitatory synapses. 

To leading order, the distribution of eigenvalues of $J$ will depend only on the mean and variance of its elements, which are summarized in the deterministic matrices $Q$ (means) and $G_{N}^{(2)}$ (variances) with elements
\begin{eqnarray}
Q_{ij} & = & P_{ij}W_{ij}\\
\left[G_{N}^{(2)}\right]_{ij} & = & P_{ij}\left(1-P_{ij}\right)W_{ij}^2
\end{eqnarray}

We will show that $\text{rank}\left\{ Q\right\} \le 3$ (generically for large $N$ and non-fine tuned parameters $\text{rank}\left\{ Q\right\} = 3$). In \cite{Tao2013}, Tao studied the spectrum of the sum of a random matrix with independent and identically distributed elements and a low-rank perturbation. The outlying eigenvalues of such a matrix fluctuate around the non-zero eigenvalues of the low-rank perturbation provided that they are outside of the bulk spectrum originating from the random part. A modification of the arguments in \cite{Tao2013} can be used to show that the same is true for the sum of a random matrix with independent but not identically distributed elements and a low-rank perturbation. 

\begin{figure}
\begin{centering}
    \includegraphics[width=1\columnwidth]{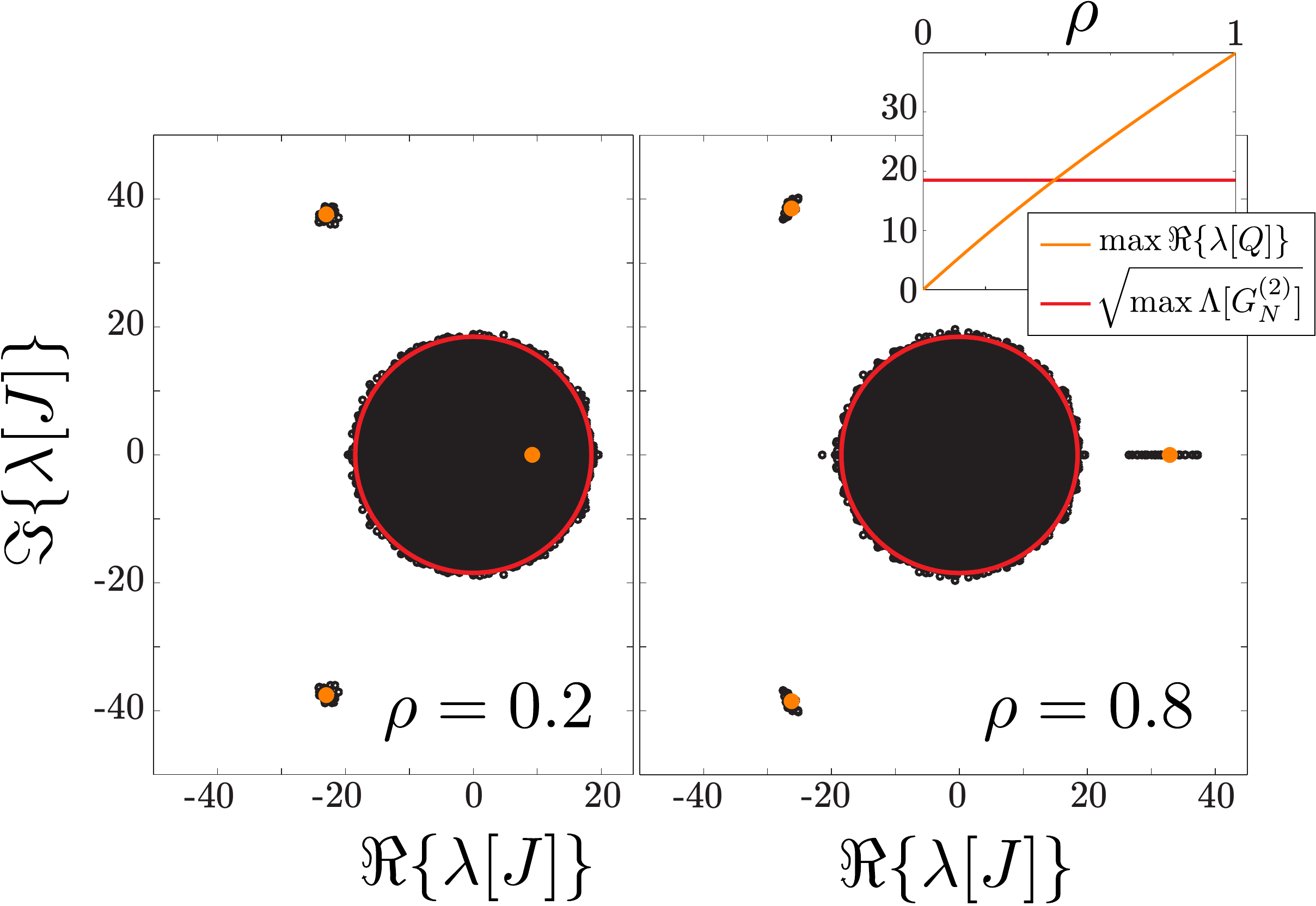}
    \caption{Spectrum of connectivity matrices with heterogeneous, correlated joint degree distribution. The network parameters were chosen to be $\kappa=0.7,\theta=28.57, N_E = 1000, N_I=250, p_0 = 0.05, W_0 = 5$, where $\kappa$ and $\theta$ are the form and scale parameters respectively of the $\Gamma$ distribution from which the in- and out-degree sequences are randomly drawn. The average correlation $\rho$ between the in- and out-degree sequences was varied between 0 and 1. For the values $\rho=0.2$ (left) and 
    $\rho=0.8$ (right) we drew 25 degree sequences and based on them drew the connectivity matrix according to
    the prescription outlined in Section \ref{sec:degdist}. The eigenvalues of each matrix were computed 
    numerically and are shown in black. For each value of $\rho$ we computed the average functions $\langle
    \mathcal{T} \rangle,~\langle \mathcal{S} \rangle$ etc. and the roots of the characteristic polynomials $
    \mathcal{A}(\Lambda)$ and $\mathcal{B}(\lambda)$  (see Appendices \ref{sec:cp} and \ref{sec:gamma} for 
    derivation). The predictions for the support of the bulk (red) and the outliers (orange) are in agreement 
    with the numerical calculation. Inset: as a function of $\rho$, there is a positive outlier that exits the 
    disk to the right.}
    \label{fig:degdist}
\end{centering}
\end{figure}

Combining these, we expect that if the non-zero eigenvalues of $Q$ are outside of the bulk that originates from the random part, the spectrum of the matrix $J$ (with non-zero means) will be approximately a composition of the bulk and outliers that can be computed separately and that the approximation will become exact as $N\to\infty$. This is verified through numerical calculations (Fig. \ref{fig:degdist}). 

Viewing the normalized degree sequences $x,y$ as deterministic variables we define
\begin{equation}
\begin{array}{rlcrl}
\mathcal{U}&=\sum_{i=1}^{N_E}x_i^2   & , & \mathcal{S}&=\sum_{i=1}^{N_E}x_i \\
           & = \sum_{i=1}^{N_E}y_i^2&,& &=\sum_{i=1}^{N_E}y_i\\ 
\mathcal{T}&=\sum_{i=1}^{N_E}x_iy_i & , & \mathcal{V}&=\sum_{i=1}^{N_E}\left(x_iy_i^2+x_i^2y_i\right)\\
\mathcal{Z}&=\sum_{i=1}^{N_E}x_i^2y_i^2 & , & \mathcal{R}&=\left(\sum_{i=1}^{N_E}x_iy_i^2\right)\left(\sum_{i=1}^{N_E}x_i^2y_i\right)
\end{array}\label{eq:degseqfunctions}
\end{equation}

Given the parameters $W_0,p_0,N_E,N_I$, we show in Appendix \ref{sec:cp} that $\text{rank}\{ G_N^{(2)}\} \le 4$ (generically for large $N$ and non-fine tuned parameters $\text{rank}\{ G_N^{(2)}\} = 4$) and its characteristic polynomial is $\mathcal{A}(\Lambda)=\sum_{k=0}^{N}\left(-1\right)^ka_k\Lambda^{N-k}$
with
\begin{widetext}
\begin{eqnarray}
a_0 & = & 1 \nonumber \\
a_1 & = & \mathcal{T}-\mathcal{Z}+N_IW_0^2p_0\left(1-p_0\right)\nonumber\\
a_2 & = & \mathcal{R}-\mathcal{ZT}+N_IW_0^2p_0\left(1-p_0\right)\left[\mathcal{T}-\mathcal{Z}-p_0\left(1-p_0\right)N_E\right]\nonumber\\
a_3 & = & N_IW_0^2p_0\left(1-p_0\right)\left\{ \mathcal{R}-\mathcal{ZT}+p_0\left(1-p_0\right)\left[\mathcal{S}^2-\mathcal{U}^2-N_E\left(\mathcal{T}-\mathcal{Z}\right)\right]\right\} \nonumber\\
a_4 & = & N_IW_0^2p_0^2\left(1-p_0\right)^2\left[N_E\left(\mathcal{ZT}-\mathcal{R}\right)-\mathcal{Z}\mathcal{S}^2-\mathcal{U}^2\mathcal{T}+\mathcal{SUV}\right],\label{eq:cpG}
\end{eqnarray}
\end{widetext}
and $a_k=0$ for $k>4$. Therefore, using our results, the radius of the bulk spectrum of $J$ is equal to the square-root of the largest solution to $\mathcal{A}(\Lambda)=0$.

Furthermore we show that the non-zero eigenvalues of $Q$ are equal to the roots of the polynomial $\mathcal{B}(\lambda)=\sum_{k=0}^{N}\left(-1\right)^kb_k\lambda^{N-k}$, with 
\begin{eqnarray}
b_0 & = & 1\nonumber\\
b_1 & = & \mathcal{T}-N_IW_0p_0\nonumber\\
b_2 & = & N_IW_0p_0\left[N_Ep_0-\mathcal{T}\right]\nonumber\\
b_3 & = & N_IW_0p_0^2\left[N_E\mathcal{T}-\mathcal{S}^2\right],\label{eq:cpQ}
\end{eqnarray}
and $b_k=0$ for $k>3$, such that the outlying eigenvalues of $J$ are approximated by the roots of $\mathcal{B}\left(\lambda\right)$ that lie outside of the bulk.

If the degree sequences are not specified, but only the joint in- and out-degree distribution they are drawn from, the random matrix $J$ will be constructed in two steps: first $k^{\rm in}$ and $k^{\rm out}$ are drawn from their joint in- and out-degree distribution, and then the elements of $J$ are drawn using the prescription outlined above. In such cases, one can in principle compute the averages $\langle \mathcal{T} \rangle, ~\langle \mathcal{S} \rangle$, etc., in terms of the moments of the joint degree distribution, and substitute these averages into the formulae we give assuming the degree sequences are fixed. 

We have carried out that calculation (Appendix \ref{sec:gamma}) for $\Gamma$ degree distributions with form parameter $\kappa$, scale parameter $\theta$ and arbitrary correlation $\rho$ of the in- and out-degree sequences (see Fig. \ref{fig:degdist}). We find that, for fixed marginals, the radius of the bulk spectrum depends extremely weakly on the correlation of the in- and out-degree sequences (see red line in inset to Fig. \ref{fig:degdist}). The matrix $Q$ however has a real, positive eigenvalue that for typical examples increases monotonically with the correlation, such that for some value it exits the bulk to the right (see Fig. \ref{fig:degdist}). Work by Roxin \cite{Roxin2011}, Schmeltzer et al. \cite{Schmeltzer2015} and unpublished work by Landau and Sompolinsky \cite{Landau2015} has shown that the broadness and correlation of the joint degree distribution can lead to qualitative changes in the behavior of a spiking network. Further work is required to investigate whether and why these changes can be explained by the spectrum of the connectivity matrix derived here.

\section{An example from ecology}\label{sec:ecology}

The past few years have seen a resurgence of interest in the use of methods from random matrix theory to study the stability of ecosystems \cite{Allesina2015a,Namba2015,Tokita2015}. While the original work by Robert May assumed a random unstructured connectivity pattern between species \cite{May1972}, experimental data shows marked departures from random connectivity \cite{Cohen1990}. This includes  hierarchical organization within ecosystems where larger species have asymmetric effect on smaller species, larger variance in the number of partners for a given species \cite{Dunne2002}, and fewer cycles  involving three or more interacting species than would be expected from an Erd\H{o}s-R\'{e}nyi graph \cite{Allesina2008}. 
A popular model for food web structure is the cascade model \cite{Cohen1985}, where species are rank-ordered, and each species can exclusively prey upon lower-ranked species. The differential effects between predators and prey in the cascade model can be described using connectivity matrices with different statistics for entries above and below the diagonal \cite{Allesina2015}:
\begin{equation}
J_{ij} = \mu(z_i,z_j)/\sqrt{N} + g(z_i,z_j)J^0_{ij}
\end{equation}
with
\begin{eqnarray}
\mu(z_i,z_j) &= &\mu_a\Theta(z_i-z_j)-\mu_b\Theta(z_j-z_i) \\
g(z_i,z_j) &=& g_a\Theta(z_i-z_j)+g_b\Theta(z_j-z_i)
\end{eqnarray}
where $\Theta$ is the Heaviside step function. We use the convention $\Theta(0)=0$. Here, $J$ describes the interactions between different species in the ecosystem. For $\mu_a,\mu_b>0$ and sufficiently larger than $g_a,g_b$, the entries above (below) the diagonal are positive (negative), so the matrix describes a perfectly hierarchical food web, where the top-ranked species consumes all the other species, the second species consumes all the species but the first, and so on.

We will focus on the random part of the matrix (i.e. we set $\mu_a=\mu_b=0$). The spectrum of the sum of the deterministic and random parts remains a problem for future study. Note that since the deterministic part has full rank, one cannot apply simple perturbation methods.

According to our analysis, the support of the spectrum of $J$ is a disk with radius $\sqrt{\Lambda_1}$, $\Lambda_1 = \lim_{N\to\infty} \max \lambda[G^{(2)}_N]$, and
\begin{equation}
G^{(2)}_N(z_i,z_j) =  N^{-1}\left(g_a^2\Theta(z_i-z_j)+g_b^2\Theta(z_j-z_i)\right)
\end{equation}
Following the derivation in \cite{Allesina2015} we will show that $\Lambda_1 = \frac{g_a^2-g_b^2}{2\log(g_a/g_b)}$.

The characteristic polynomial $\mathcal{D}_N(\lambda) = \det|I\lambda-G_N^{(2)}|$ is simplified by subtracting the $i+1$ column from the $i$-th column for $i=1,\dots,N-1$ giving
\begin{align}
\mathcal{D}_N(\lambda) = {\rm det}\left|\begin{matrix}
                                            \lambda + a   & 0      & 0            & -a      \\
                                            -(\lambda+b)  & \ddots & 0            & \vdots  \\
                                            0             & \ddots & \lambda+a    & -a      \\
                                            0             & 0      & -(\lambda+b) & \lambda
                                        \end{matrix}\right|.
\end{align}
Where we have defined $a = g_a^2/N$  and $b = g_b^2/N$.
This simplifies to the recursion relation $\mathcal{D}_N(\lambda) = (\lambda + a)\mathcal{D}_{N-1}(\lambda) -a(\lambda + b)^{N-1}$. Taking into account that $\mathcal{D}_2(\lambda)=\lambda^2-ab$, this recursion relation can be solved, giving:
\begin{eqnarray}
\mathcal{D}_N(\lambda) & = & \frac{1}{a-b}\left[a\left(\lambda+b\right)^{N}-b\left(\lambda+a\right)^{N}\right].
\end{eqnarray}
Setting the characteristic polynomial $\mathcal{D}_N(\lambda)$ to 0 leads to the equation
\begin{equation}
b=a \left(\frac{b+\lambda}{a+\lambda}\right)^{N}
\end{equation}
which has multiple roots
\begin{equation}
\lambda_k = a\frac{\left(\frac{b}{a}\right)^{\frac{1}{N}}e^{\frac{2\pi i k}{N}}-\frac{b}{a}}{1-\left(\frac{b}{a}\right)^{\frac{1}{N}}e^{\frac{2\pi i k}{N}}},\qquad k=1,\dots,N.
\end{equation}
We are interested in the largest among the $N$ roots, which is real and positive. Taking into account the dependence of $a$ and $b$ on $N$, we find that:
\begin{equation}
\Lambda_1 = \lim_{N\to\infty} \max_k \left[ a\frac{\left(\frac{b}{a}\right)^{\frac{1}{N}}e^{\frac{2\pi i k}{N}}-\frac{b}{a}}{1-\left(\frac{b}{a}\right)^{\frac{1}{N}}e^{\frac{2\pi i k}{N}}}\right] = \frac{g_a^2-g_b^2}{\log\left(\frac{g_a^2}{g_b^2}\right)},
\end{equation}
as desired.

Interestingly, for all values of $g_a,g_b$ the spectral radius of $J$ is smaller than the radius of the network if the predator-prey structure did not exist. The latter is equal to $\sqrt{(g_a^2+g_b^2)/2}$. This suggests that the hierarchical structure of the interaction network serves to stabilize the ecosystem regardless of how dominant the predators are over the prey.

Note however that in this model there are no correlations. In \cite{Allesina2015}, it was shown numerically that correlations (i.e. the expectation value of $J_{ij}^0J_{ji}^0$) can dramatically change the stability of the network, compared with one that has no correlations.

\section{Discussion and conclusions}\label{sec:conc}

We studied jointly the spectrum of a new random matrix model and the dynamics of the neural network model it implies. We found that, as a function of the deterministic structure of the network (given by $g$), the network becomes spontaneously active at a critical point. 

Identifying a space where the dynamics of a neural network can be described efficiently and robustly is one of the challenges of modern neuroscience \cite{Gao2015}. In our model, above the critical point, the deterministic dynamics of the entire network are well approximated by a potentially low dimensional probability distribution, with dimension equal to the number of eigenvalues of a deterministic matrix that have real part greater than 1. 

Two limitations of using the results of our previous studies \cite{Aljadeff2014a,Aljadeff2015} to interpret multi-unit recordings are that it requires knowing the cell-type identity of each neuron in the network and it only provides a prediction for quantities averaged over all neurons of a specific type.

Here these are remedied. First, while some information about the connectivity structure is still required, this could be in the form of global spatial symmetries (``rules'') present in the network, such as the connectivity rule we used in the ring model.
Second, our analysis provides a prediction for single neuron quantities, namely the participation of every neuron in the network in the global active dynamic modes.

Existence of discrete network modules with no apparent fine-tuned connectivity has been shown to exist in networks of grid-cells in mammalian medial entorhinal cortex \cite{Stensola2012}. These cells fire when the animal's position is on the vertices of a hexagonal lattice, and are thought to be important for spatial navigation. Interestingly, when characterizing the firing properties of many such cells in a single animal one finds that the the lattice spacing of all cells belongs (approximately) to a discrete set that forms a geometric series \cite{Stensola2012}. Much work has been devoted to trying to understand how such a code could be used efficiently to represent the animal's location (see for example \cite{Stemmler2015,Wei2015}) and how such a code could be generated \cite{Burak2009}.

However, we are not aware of a model that explains how multiple modules (sub-networks with distinct grid spacing) could be generated without fine-tuned connectivity, which is not observed experimentally. In our model, continuous changes to a connectivity parameter can introduce additional discrete and spatially periodic modes into the network represented by finer and finer lattices. We are not arguing that the random network we are studying here could serve as a model of grid-cell networks, as there are many missing details that cannot be accounted for by our model. Nevertheless our analysis uncovers a mechanism by which a low-dimensional, spatially structured dynamics could arise as a result of random connectivity.


\section*{Acknowledgments}

The authors would like to thank Nicolas Brunel, Stefano Allesina, Alex Roxin and James Heys for discussions. JA was supported by NSF CRCNS IIS-1430296. MV acknowledges a doctoral grant by Fundaci\'{o} ``la Caixa'' and a travel grant by Fundaci\'{o} Ferran Sunyer i Balaguer. TOS was supported by grants No. R01EY019493 and No. P30 EY019005 and NSF Career Award (IIS 1254123) and by the Salk Innovations program.

\appendix
\section{The limit $K,N\to\infty$}\label{sec:KN}

Here we will show that the difference between the piecewise estimate $\tilde{g}$ and the continuous synaptic gain function $g$ goes to $0$ as $N\to\infty$. We assumed that the unit square can be tiled by square subsets of area $s_0^2>0$ where $g$ is bounded, differentiable, and its first derivative is bounded in each subset. Note that the with Lipschitz constant of $g$ can depend on $N$, but $s_0$ cannot. 

For $N,K(N)\in\mathbb{N}$, recall our definitions for $\tilde{g}$ and $\mu_i$ (Eqs. \ref{eq:mui}, \ref{eq:piecewise}) and  define $k_{ij} = \left( K^{-1} (\mu_i-1) , K^{-1} \mu_i\right]\times\left( K^{-1} (\mu_j-1) , K^{-1} \mu_j\right]$. Also recall our assumption  each point is either inside a square with side $s_0$ within which there are no discontinuities or on the border of such a subset. Thus, for $K>s_0^{-1}$ we can assume that every constant region of $\tilde{g}$ is contained within a single square subset.

We would like to show that for all $i,j$
\begin{equation}
\lim_{N\to\infty} |\tilde{g}_N(z_i,z_j) -g(z_i,z_j)| = 0. \label{eq:limN} 
\end{equation}
Since $s_0$ is independent of $N$, we only have to show that Eq. (\ref{eq:limN}) is true within a subset where $g$ satisfies the smoothness conditions.

Using our definitions and the fact that $g$ has Lipschitz constant $C_L(N) = C_L^0 N^\beta$,
\begin{widetext}
\begin{eqnarray}
\left|\tilde{g}_N(z_i,z_j) -g(z_i,z_j)\right| & = & \left|g\left(\frac{\mu_i-\frac{1}{2}}{K},\frac{\mu_j-\frac{1}{2}}{K}\right) -g(z_i,z_j)\right| \le 	\sup_{(z^\prime_i,z^\prime_j)\in k_{ij}} \left|g\left(\frac{\mu_i-\frac{1}{2}}{K},\frac{\mu_j-\frac{1}{2}}{K}\right) -g(z^\prime_i,z^\prime_j)\right| \nonumber\\
 & \le & C_L\sup_{(z^\prime_i,z^\prime_j)\in k_{ij}} \left[\left(\frac{\mu_i-\frac{1}{2}}{K}-z^\prime_i\right)^2 + \left(\frac{\mu_j-\frac{1}{2}}{K}-z^\prime_j\right)^2\right]^{\frac{1}{2}} =  C^0_L\frac{N^{\beta}}{2K}
\end{eqnarray}
\end{widetext}
So finally, 
\begin{equation}
\lim_{N\to\infty} |\tilde{g}_N(z_i,z_j) -g(z_i,z_j)| \le \frac{C^0_L}{2} \lim_{N\to\infty}\frac{N^{\beta}}{K(N)} = 0.
\end{equation}


\section{The characteristic polynomials of $G_N^{(2)}$ and $Q$.}\label{sec:cp}

Here we compute directly the characteristic polynomials of $G_N^{(2)}$ and $Q$ (Eqs. \ref{eq:cpG}, \ref{eq:cpQ}) using the minor expansion formula.

\subsection{Calculation of spectrum of $G^{(2)}$}

Recall that $N=N_E+N_I$, and let $\mathcal{G}^{(2)}_{k}$ be the $k\times k$ matrix with elements taken from the intersection of $k$  specific rows and columns of $G^{(2)}_N$. The notation $\mathcal{G}^{(2)}_{k_E,k_I}$ will indicate that exactly $k_E$ and $k_I$ of these rows and columns correspond to excitatory and inhibitory neurons, respectively. 

For convenience we will use $v=p_0\left(1-p_0\right)$ and $w=W_0^2p_0\left(1-p_0\right)$. We would like to write an expression for the characteristic polynomial of $G_{N_E+N_I}^{(2)}$ using the sums over its diagonal minors
\begin{eqnarray}
\mathcal{A}_{N_E,N_I}\left(\Lambda\right) & = & \sum_{k=0}^{N}\left(-1\right)^ka_k\Lambda^{N-k}\label{eq:appcpG}
\end{eqnarray}
where $a_k=\sum\det \mathcal{G}_k^{(2)}$ for $k\ge 1$ and $a_0=1$. The notation $\sum\det \mathcal{G}_k^{(2)}$ means a sum over all combinations of $N_E,N_I$ such that $N_E+N_I=k$ (i.e. the so-called $k$-row diagonal minors of $G_{N}^{(2)}$). We will compute
$a_0,\dots,a_4$ explicitly and show that $a_k=0$ for $k>4$. 

We begin by noting that the determinant of the $3\times3$ matrix $\mathcal{G}_{3,0}^{(2)}  =  {\rm diag}\left(x_1,x_2,x_3\right) \left(\begin{smallmatrix}
1-x_1y_1 & 1-x_1y_2 & 1-x_1y_3\\
1-x_2y_1 & 1-x_2y_2 & 1-x_2y_3\\
1-x_3y_1 & 1-x_3y_2 & 1-x_3y_3
\end{smallmatrix}\right){\rm diag}\left(y_1,y_2,y_3\right)$ is 0 because the middle matrix is the sum of two rank 1 matrices.

\subsubsection*{$a_0$.}
By definition, $a_0=1$. 

\subsubsection*{$a_1$.}
The second coefficient, $a_1$ is simply the trace 
\begin{eqnarray}
\text{Tr}\left\{ G_{N_E,N_I}^{(2)}\right\}  & = & \sum_{i=1}^{N_E}x_iy_i\left(1-x_iy_i\right)+N_Iw \nonumber \\
a_1 & = & \mathcal{T}-\mathcal{Z}+N_Iw,
\end{eqnarray}
where in the second row we used the functions of the degree sequences (Eq. \ref{eq:degseqfunctions}).
\begin{widetext}
\subsubsection*{$a_2$.}
The third coefficient $a_2$ is the sum of $2$ row diagonal minors. There are three types of diagonal minors, only two of which are non-zero 
\begin{eqnarray}
\det \mathcal{G}_{2,0}^{(2)} & = & \det\left(\begin{array}{cc}
x_i & 0\\
0     & x_j
\end{array}\right)\det\left(\begin{array}{cc}
\left(1-x_iy_i\right) & \left(1-x_iy_j\right)\\
\left(1-x_jy_i\right) & \left(1-x_jy_j\right)
\end{array}\right)\det\left(\begin{array}{cc}
y_i & 0 \\
0     & y_j
\end{array}\right) = x_ix_jy_iy_j\left(x_iy_j+x_jy_i-x_iy_i-x_jy_j\right)\nonumber\\
\det \mathcal{G}_{1,1}^{(2)} & = & \det\left(\begin{array}{cc}
x_iy_i\left(1-x_iy_i\right) & w\\
v & w
\end{array}\right) =  w\left[x_iy_i\left(1-x_iy_i\right)-v\right]\nonumber\\
\det \mathcal{G}_{0,2}^{(2)} & = & \det\left(\begin{array}{cc}
w & w\\
w & w
\end{array}\right)=0
\end{eqnarray}
Carrying out the summation over possible combinations 
\begin{eqnarray}
\sum\det \mathcal{G}_{2,0}^{(2)} & = & \sum_{i<j}x_ix_jy_iy_j\left(x_iy_j+x_jy_i-x_iy_i-x_jy_j\right)= \frac{1}{2}\sum_{i=1}^{N_E}\sum_{j=1}^{N_E}x_ix_jy_iy_j\left(x_iy_j+x_jy_i-x_iy_i-x_jy_j\right)\nonumber\\
 & = & \mathcal{R}-\mathcal{ZT}\nonumber\\
\sum\det \mathcal{G}_{1,1}^{(2)} & = & N_Iw\sum_{i=1}^{N_E}\left[x_iy_i\left(1-x_iy_i\right)-v\right]= N_Iw\left[\mathcal{T}-\mathcal{Z}-v N_E\right]
\end{eqnarray}
Putting these together we get 
\begin{equation}
a_2 = \mathcal{R}-\mathcal{ZT}+N_Iw\left[\mathcal{T}-\mathcal{Z}-vN_E\right].
\end{equation}

\subsubsection*{$a_3$.}
The fourth coefficient $a_3$ is the sum of all $3$ row diagonal minors. Now there are four types of minors, only one of which is non-zero
\begin{eqnarray}
\det \mathcal{G}_{3,0}^{(2)} & = & 0\qquad\text{(shown above)}\nonumber\\
\det \mathcal{G}_{2,1}^{(2)} & = & \det\left(\begin{array}{ccc}
x_iy_i\left(1-x_iy_i\right) & x_iy_j\left(1-x_iy_j\right) & w\\
x_jy_i\left(1-x_jy_i\right) & x_jy_j\left(1-x_jy_j\right) & w\\
v & v & w
\end{array}\right) \nonumber\\
& =&  w\det \mathcal{G}_{2,0}^{(2)}+vw\left[x_jy_i\left(1-x_jy_i\right)+x_iy_j\left(1-x_iy_j\right)-x_iy_i\left(1-x_iy_i\right)-x_jy_j\left(1-x_jy_j\right)\right]\nonumber\\
\det \mathcal{G}_{1,2}^{(2)} & = & \det \mathcal{G}_{0,3}^{(2)}=0\qquad\text{(repeated columns of inhibitory neurons)}
\end{eqnarray}
Carrying out the sum
\begin{eqnarray}
a_3 & = & \sum\det \mathcal{G}_{2,1}^{(2)}\nonumber\\
 & = & w N_I\left[\mathcal{R}-\mathcal{ZT}\right]+ vw N_I\frac{1}{2}\sum_{i=1}^{N_E}\sum_{j=1}^{N_E}\left[x_jy_i\left(1-x_jy_i\right)+x_iy_j\left(1-x_iy_j\right)-x_iy_i\left(1-x_iy_i\right)-x_jy_j\left(1-x_jy_j\right)\right]\nonumber\\
 & = & w N_I\left\{ \mathcal{R}-\mathcal{ZT}+v\left[\mathcal{S}^2-\mathcal{U}^2-N_E\left(\mathcal{T}-\mathcal{Z}\right)\right]\right\} 
\end{eqnarray}
\subsubsection*{$a_4$.} 
The last non-zero coefficient is $a_4$, the sum of all $4$ row diagonal minors. Here there are five types, only one of which is non-zero: 
\begin{eqnarray}
\det \mathcal{G}_{4,0}^{(2)} & = & 0\qquad\text{(because \ensuremath{\det \mathcal{G}_{3,0}^{(2)}=0})}\nonumber\\
\det \mathcal{G}_{3,1}^{(2)} & = & \det\left(\begin{array}{cccc}
x_iy_i\left(1-x_iy_i\right) & x_iy_j\left(1-x_iy_j\right) & x_iy_k\left(1-x_iy_k\right) & w\\
x_jy_i\left(1-x_jy_i\right) & x_jy_j\left(1-x_jy_j\right) & x_jy_k\left(1-x_jy_k\right) & w\\
x_ky_i\left(1-x_ky_i\right) & x_ky_j\left(1-x_ky_j\right) & x_ky_k\left(1-x_ky_k\right) & w\\
v & v & v & w
\end{array}\right)\nonumber\\
 & = & vw\left(x_i-x_j\right)\left(x_i-x_k\right)\left(x_j-x_k\right)\left(y_i-y_j\right)\left(y_i-y_k\right)\left(y_j-y_k\right)\nonumber\\
\det \mathcal{G}_{2,2}^{(2)} & = & \det \mathcal{G}_{1,3}^{(2)}=\det \mathcal{G}_{0,4}^{(2)}=0\qquad\text{(repeated columns of inhibitory neurons)}
\end{eqnarray}
Carrying out the sum we get
\begin{eqnarray}
a_4 & = & \frac{1}{6}vw N_I\sum_{i=1}^{N_E}\sum_{j=1}^{N_E}\sum_{k=1}^{N_E}\left(x_i^2x_j-x_i^2x_k+x_j^2x_k-x_j^2x_i+x_k^2x_i-x_k^2x_j\right)\left(y_i^2y_j-y_i^2y_k+y_j^2y_k-y_j^2y_i+y_k^2y_i-y_k^2y_j\right)\nonumber\\
 & = & N_Ivw\left[N_E\left(\mathcal{ZT}-\mathcal{R}\right)-\mathcal{ZS}^2-\mathcal{U}^2\mathcal{T}+\mathcal{SUV}\right]
\end{eqnarray}
\end{widetext}

\subsubsection*{$a_k,~k>4$.}
Now we show that $a_k=0$ for $k>4$. A diagonal minor representing a subnetwork of five neurons or more can have $N_I=0$, $N_I=1$, or $N_I\ge 2$. If $N_I\ge2$ the diagonal minor is zero because of repeated columns. If $N_I=1$, then $N_E\ge4$. Here, the determinant is a weighted sum of $k=N_E-1=N-2$ row diagonal minors of the form $\det \mathcal{G}_{N_E-1,0}^{(2)}$ which is zero for $N_E\ge3$. Lastly if $N_I=0$ then again we have a sum of terms of the form $\det \mathcal{G}_{N_E,0}^{(2)}$ which are zero as discussed above.

\subsection{Calculation of spectrum of $Q$}

Using a similar approach we will compute the characteristic polynomial of $Q$ and show that generically $\text{rank}\left\{ Q\right\} =3$. Using the sums over diagonal minors of $Q_{N_E+N_I}$ 
\begin{equation}
\mathcal{B}_{N_E,N_I}\left(\lambda\right) = \sum_{k=0}^{N}\left(-1\right)^kb_k\lambda^{N-k}
\end{equation}
where $b_k=\sum\det \mathcal{Q}_k$ for $k\ge 1$ and where $\mathcal{Q}_k$ is a $k\times k$ matrix with elements taken from the intersection of $k$ rows and columns of $Q$. Again, $\mathcal{Q}_{k_E,k_I}$ will indicate that $k_E$ and $k_I$ rows and columns correspond to excitatory and inhibitory neurons, respectively. 

\subsubsection*{$b_0$.}
By definition we have $b_0=1$. 
\subsubsection*{$b_1$.}
The second term is the trace
\begin{eqnarray}
b_1 & = & \text{Tr}\left\{ Q_{N_E+N_I}\right\}  = \sum_{i=1}^{N_E}x_iy_i-N_IW_0p_0\nonumber\\
 & = & \mathcal{T}-N_IW_0p_0
\end{eqnarray}

\subsubsection*{$b_2$.}
The third coefficient is the sum over $2$ row diagonal minors
\begin{eqnarray}
\det \mathcal{Q}_{2,0} & = & \det\left(\begin{array}{cc}
x_iy_i & x_iy_j\\
x_jy_i & x_jy_j
\end{array}\right)=0\nonumber \\
\det \mathcal{Q}_{1,1} & = & \det\left(\begin{array}{cc}
x_iy_i & -p_0W_0\\
p_0 & -p_0W_0
\end{array}\right) =  p_0W_0\left(p_0-x_iy_i\right)\nonumber\\
\det \mathcal{Q}_{0,2} & = & \det\left(\begin{array}{cc}
-p_0W_0 & -p_0W_0\\
-p_0W_0 & -p_0W_0
\end{array}\right)=0
\end{eqnarray}
carrying out the summation, we get
\begin{eqnarray}
b_2 & = & p_0W_0N_I\sum_{i=1}^{N_E}\left(p_0-x_iy_i\right) \nonumber \\ 
& = &  p_0W_0N_I\left(N_Ep_0-\mathcal{T}\right)
\end{eqnarray}
\subsubsection*{$b_3$.}
The fourth and last non-zero coefficient is the sum over $3$ row diagonal minors 
\begin{eqnarray}
\det \mathcal{Q}_{3,0} & = & 0\nonumber\\
\det \mathcal{Q}_{2,1} & = & \left(\begin{array}{ccc}
x_iy_i & x_iy_j & -p_0W_0\\
x_jy_i & x_jy_j & -p_0W_0\\
p_0 & p_0 & -p_0W_0
\end{array}\right)\nonumber\\
 & = & p_0^2W_0\left(x_iy_i+x_jy_j-x_iy_j-x_jy_i\right)\nonumber\\
\det \mathcal{Q}_{1,2} & = & \det Q_{0,3}=0\quad\text{(repeated columns)}
\end{eqnarray}
Carrying out the sum 
\begin{eqnarray}
b_3 & = & N_Ip_0^2W_0\sum_{i<j}\left(x_iy_i+x_jy_j-x_iy_j-x_jy_i\right)\nonumber\\
 & = & \frac{1}{2}N_Ip_0^2W_0\sum_{i=1}^{N_E}\sum_{j=1}^{N_E}\left(x_iy_i+x_jy_j-x_iy_j-x_jy_i\right)\nonumber\\
 & = & N_Ip_0^2W_0\left(N_E\mathcal{T}-\mathcal{S}^2\right)
\end{eqnarray}

\subsubsection*{$b_k,~k>3$.}
Now we show that $b_k=0$ for $k>3$. A minor representing a subnetwork of four neurons or more can have $N_I=0$, $N_I=1$, or $N_I\ge2$. If $N_I\ge2$ the minor is zero because of repeated columns. If $N_I=1$, then $N_E\ge3$. Here, the determinant is a sum of $k=N_E-1=N-2$ row diagonal minors of the form $\det \mathcal{Q}_{N_E-1,0}$ which is zero for $N_E\ge2$. Lastly if $N_I=0$ then again we have a sum of terms of the form $\det \mathcal{Q}_{N_E,0}$ which are zero as discussed above. 

\section{Networks with $\Gamma$ degree distributions}\label{sec:gamma}

We choose a specific parameterization where the marginals of the joint in- and out-degree distribution are $\Gamma$ with form parameter $\kappa$, scale parameter $\theta$ and have average correlation $\rho$. Owing to the properties of sums of random variables that follow a $\Gamma$ distribution, we can write the random in- and out-degree sequences as
\begin{align}
k_i^{\text{in}} & = & k_{1i}+k_{2i}  & \quad, &  k_{1i} & \sim  \Gamma\left(\kappa\rho,\theta\right)\nonumber \\
k_i^{\text{out}} & = & k_{1i}+k_{3i} & \quad, & k_{2i},k_{3i} & \sim \Gamma\left(\kappa\left(1-\rho\right),\theta\right)
\end{align}
where $1\le i \le N_E$. In this Appendix $\langle\cdot\rangle$ will denote averages over the joint in- and out-degree distribution. 

The moments of the $\Gamma$ distribution imply that, for this parametrization,
\begin{equation}
\langle (k_i^{\text{in}})^{n}\rangle  = \langle (k_i^{\text{out}})^{n}\rangle =\theta^{n}\prod_{m=0}^{n-1}(\kappa+m).
\end{equation}
for all $1\le i \le N_E$. Here, since elements of $k_i^{\text{in}}$ and $k_i^{\text{in}}$ are (separately) independent and identically distributed we will suppress the subscript $i$ and superscripts ${\rm in, out}$ when possible, and let $\langle k^2\rangle = \langle k^{{\rm in}\top}k^{\rm in}\rangle$, $k^{\rm in}k^{\rm out}=k^{{\rm in}\top}k^{\rm out}$ etc. 

One can verify that indeed the average correlation between the in- and out-degree sequences is
\begin{equation}
\frac{\left\langle k^{\text{in}}k^{\text{out}}\right\rangle -\left\langle k^{\text{in}}\right\rangle \left\langle k^{\text{out}}\right\rangle }{\sqrt{\left\langle k^{\text{in}2}-\left\langle k^{\text{in}}\right\rangle ^2\right\rangle }\sqrt{\left\langle k^{\text{out}2}-\left\langle k^{\text{out}}\right\rangle ^2\right\rangle }} = \rho.
\end{equation}

Using this parametrization we compute the averages $\langle\mathcal{T}\rangle,~\langle\mathcal{S}\rangle$ etc. and express them in terms of $\rho,\theta,\kappa$ and $N_E$.

\subsubsection*{$\mathcal{T}$.}

\begin{eqnarray}
\mathcal{T} & = & \sum_{i=1}^{N_E} x_iy_i=\frac{1}{N_E\kappa\theta}\sum_{i=1}^{N_E}k_i^{\text{in}}k_i^{\text{out}}\nonumber\\
\left\langle \mathcal{T}\right\rangle  & = & \frac{1}{\kappa\theta}\left\langle k^{\text{in}}k^{\text{out}}\right\rangle =\theta\left(\rho+\kappa\right)
\end{eqnarray}

\subsubsection*{$\mathcal{S}$.}

\begin{eqnarray}
\mathcal{S} & = & \sum_{i=1}^{N_E}x_i=\frac{1}{\sqrt{N_E\kappa\theta}}\sum_{i=1}^{N_E}k_i^{\text{in}}\nonumber\\
\left\langle \mathcal{S}\right\rangle  & = & \sqrt{\frac{N_E}{\kappa\theta}}\left\langle k\right\rangle =\sqrt{N_E\kappa\theta}
\end{eqnarray}

\subsubsection*{$\mathcal{U}$.}

\begin{eqnarray}
\mathcal{U} & = & \sum_{i=1}^{N_E}x_i^2=\frac{1}{N_E\kappa\theta}\sum_{i=1}^{N_E}k_i^{\text{in}2}\nonumber\\
\left\langle \mathcal{U}\right\rangle  & = & \frac{1}{\kappa\theta}\left\langle k^2\right\rangle =\theta\left(\kappa+1\right)
\end{eqnarray}

\subsubsection*{$\mathcal{Z}$.}

To compute $\langle\mathcal{Z}\rangle$ we first derive an expression for $\left\langle k^{\text{in}2}k^{\text{out}2}\right\rangle$. Using the independence of $k_1,k_2,k_3$:
\begin{eqnarray}
\left\langle k^{\text{in}2}k^{\text{out}2}\right\rangle  & = & \left\langle \left(k_1^2+2k_1k_2+k_2^2\right)\left(k_1^2+2k_1k_3+k_3^2\right)\right\rangle \nonumber\\
 & = & \theta^4\Big\{ 6\kappa\rho+\kappa^2\left[1+8\rho+2\rho^2\right]+\nonumber \\
 &  & \qquad 2\kappa^3\left[1+2\rho\right]+\kappa^4\Big\}.
\end{eqnarray}
Now we can write
\begin{eqnarray}
\mathcal{Z} & = & \sum_{i=1}^{N_E}x_i^2y_i^2=\frac{1}{N_E^2\kappa^2\theta^2}\sum_{i=1}^{N_E}k_i^{\text{in}2}k_i^{\text{out}2}\nonumber\\
\left\langle \mathcal{Z}\right\rangle  & = & \frac{1}{N_E\kappa^2\theta^2}\left\langle k^{\text{in}2}k^{\text{out}2}\right\rangle \nonumber\\
 & = & \frac{\theta^2}{N_E}\Big\{ 6\frac{\rho}{\kappa}+\left[1+8\rho+2\rho^2\right]+\nonumber \\
 &  & \qquad 2\kappa\left[1+2\rho\right]+\kappa^2\Big\} 
\end{eqnarray}

\subsubsection*{$\mathcal{R}$.}

To compute $\langle\mathcal{R}\rangle$ (and $\langle\mathcal{V}\rangle$) we first derive an expression for $\left\langle k^{\text{in}}k^{\text{out}2}\right\rangle$. Using the independence of $k_1,k_2,k_3$
\begin{eqnarray}
\left\langle k^{\text{in}2}k^{\text{out}}\right\rangle  & = & \left\langle \left(k_1^2+2k_1k_2+k_2^2\right)\left(k_1+k_3\right)\right\rangle \nonumber\\
 & = & \theta^3\kappa\left(\kappa+1\right)\left(\kappa+2\rho\right)\nonumber\\
\left\langle k^{\text{in}}k^{\text{out}2}\right\rangle  & = & \left\langle \left(k_1+k_2\right)\left(k_1^2+2k_1k_3+k_3^2\right)\right\rangle \nonumber\\
 & = & \theta^3\kappa\left(\kappa+1\right)\left(\kappa+2\rho\right)
\end{eqnarray}
Now we can write
\begin{eqnarray}
\mathcal{R} & = & \frac{1}{N_E^3\kappa^3\theta^3}\left(\sum_{i=1}^{N_E}k_i^{\text{in}}k_i^{\text{out}2}\right)\left(\sum_{i=1}^{N_E}k_i^{\text{in}2}k_i^{\text{out}}\right)\nonumber\\
\left\langle \mathcal{R}\right\rangle & = & \frac{1}{N_E\kappa^3\theta^3}\left\langle k^{\text{in}}k^{\text{out}2}\right\rangle \left\langle k^{\text{in}2}k^{\text{out}}\right\rangle \nonumber\\
  & = & \frac{\theta^3\left(\kappa+1\right)^2\left(\kappa+2\rho\right)^2}{N_E\kappa}
\end{eqnarray}

\subsubsection*{$\mathcal{V}$.}

\begin{eqnarray}
\mathcal{V} & = & \frac{1}{N_E^3\kappa^3\theta^3}\sum_{i=1}^{N_E}\left(k^{\text{in}}k^{\text{out}2}+k^{\text{in}2}k^{\text{out}}\right)\nonumber\\
\left\langle \mathcal{V}\right\rangle & =& \frac{1}{N_E\kappa^3\theta^3}\left(\left\langle k^{\text{in}}k^{\text{out}2}\right\rangle +\left\langle k^{\text{in}2}k^{\text{out}}\right\rangle \right)\nonumber\\
 & = & \frac{2\left(\kappa+1\right)\left(\kappa+2\rho\right)}{N_E\kappa^2}
\end{eqnarray}

\end{document}